\documentclass[twoside]{dis09}
\usepackage[latin1]{inputenc}
\usepackage[dvips]{graphicx,epsfig,color}
\usepackage{wrapfig,rotating}
\usepackage{amssymb,amsmath,array}

\begin{document}
\title{Exclusive vs. Diffractive Vector Meson Production in DIS at small $x$ or off Nuclei}

\author{Cyrille Marquet$^{1,2}$ and Bin Wu$^{1,3}$
%
%
\vspace{.3cm}\\
%
1- Department of Physics, Columbia University, New York, NY 10027, USA
%
\vspace{.1cm}\\
2- Institut de Physique Th\'eorique, CEA/Saclay, 91191 Gif-sur-Yvette cedex, France
\vspace{.1cm}\\
3- Department of Physics, Peking University, Beijing, 100871, P.R. China
}

\maketitle

\begin{abstract}

The diffractive production of vector mesons in deep inelastic scattering (DIS) is calculated
in the McLerran-Venugopalan model. This is relevant when large parton densities are probed by the virtual photon, as is the case at small Bjorken $x$ or in DIS off nuclei. We investigate differences between the exclusive production (when the target doesn't break up) which dominates at small momentum transfer squared $|t|,$ and the diffractive production (when the target scatters inelastically) which dominates at large $|t|.$

\end{abstract}

\section{Motivation}

Diffractive vector meson production in deep inelastic scattering (DIS) $\gamma^* A \to V Y,$ where $A$ stands for the target nucleus and $Y$ for the final state it may dissociate into, is a process in which elastic and inelastic interactions of the target can be experimentally distinguished. At high energies, the $q\bar q$ dipole that the virtual photon has fluctuated into scatters off the gluonic field of the nucleus before recombining into the vector meson. While this scattering involves a color-singlet exchange, leaving a rapidity gap in the final state, the nucleus can still interact elastically ($Y=A,$ this is called coherent diffraction) or inelastically ({\it i.e.} break up, called incoherent diffraction).

Kinematically, a low invariant mass of the system $Y$ corresponds to a large rapidity gap in the final state between that system and the vector meson, and implies that the longitudinal momentum of the meson is close to that of the incoming photon. In this case, the eikonal approximation can be assumed to compute the dipole-nucleus scattering. At small values of $x=(Q^2+M_V^2)/(Q^2+W^2)$ where $Q^2$ is the photon virtuality, $M_V$ the vector meson mass, and $W$ the energy of the
$\gamma^*-A$ collision, a target proton can also be considered. Indeed in that case, since partons with an energy fraction as small as $x$ are probed in the target wave function, the dipole will
scatter off large gluon densities generated by the QCD evolution.

The cross-section is maximal at minimum momentum transfer with exclusive production (or coherent diffraction) dominating. As the transfer of momentum gets larger, the role of incoherent diffraction increases and eventually it becomes dominant, typically for momenta bigger that the inverse target size; the elastic contribution decreases exponentially while the inelastic contribution decreases only as a power law. In the case of a target proton, it is known that saturation models describe well the exclusive cross section \cite{vm-sat}, while the BFKL Pomeron exchange approach works well for the target-dissociation cross-section \cite{vm-bfkl}.

In this work we show that, with the Color Glass Condensate (CGC) picture of the small$-x$ part of the hadronic/nuclear wave function, both coherent and incoherent diffraction can be described in the same framework, for protons and nuclei. We also explicitly calculate both contributions to
the diffractive vector meson production cross-section using the McLerran-Venugopalan (MV) model for the CGC wave function. Finally, we discuss phenomenological consequences in the context of a future electron-ion collider \cite{eic}.

\section{Diffractive Vector Meson Production in DIS off the CGC}

The CGC is an effective theory of QCD \cite{cgcrev} which aims at describing the small $x$ part of the hadronic/nuclear wave function, when the gluon density is so large that non-linear effects are important. Rather than using a standard Fock-state decomposition, it is more efficient to describe it with collective degrees of freedom, more adapted to account for the collective behavior of the small-$x$ gluons, which do not interact with a probe independently, but rather behave coherently. The CGC approach uses classical color fields: 
\begin{equation}
|h\rangle=|qqq\rangle+|qqqg\rangle+\dots+|qqqg\dots ggg\rangle+\dots\quad
\Rightarrow\quad|h\rangle=\int D{\cal A}\ \Phi_{x_A}[{\cal A}]\ |{\cal A}\rangle
\label{cgc}\ .\end{equation}
The long-lived, large-$x$ partons are represented by a strong color source $\rho$ which is static during the lifetime of the short-lived small-$x$ gluons, whose dynamics is described by the color field ${\cal A}\!\sim\!1/g_S.$ $x_A$ denotes an arbitrary separation between the field and the source.

The CGC wavefunction $\Phi_{x_A}[{\cal A}]$ is the fundamental object of this picture, it is mainly a non-perturbative quantity, but the $x_A$ evolution can be computed perturbatively. Requiring that observables are independent of the choice of $x_A,$ a functional renormalization group equation can be derived. In the leading-logarithmic approximation which resums powers of $\alpha_S\ln(1/x_A),$ the JIMWLK equation \cite{jimwlk} describes the evolution of $|\Phi_{x_A}[{\cal A}]|^2$ with $x_A.$ The information contained in the wavefunction, on gluon number and gluon correlations, can be expressed in terms of n-point correlators, probed in scattering processes. These correlators consist of Wilson lines averaged with the CGC wavefunction, and resum powers of $g_S{\cal A}.$

In diffractive vector meson production, the relevant quantity is (the photon is a right mover, the CGC a left mover, and the gauge is ${\cal A}^+=0$):
\begin{equation}
T_{\textbf{xy}}[{\cal A}^-]=1-\frac{1}{N_c}
\mbox{Tr}\left(U^\dagger_{\textbf{y}}U_{\textbf{x}}\right)\ ,\quad\mbox{with }
U_{\textbf{x}}[{\cal A}^-]={\cal P}
\exp\left(ig_S\int dz^+ T^c {\cal A}_c^-(z^+,\textbf{x})\right)\ .
\label{wline}\end{equation}
In terms of this object, the differential cross sections for a transversely (T) or longitudinally (L) polarized photon are given by (with $t=-q_\perp^2$ the momentum transfer squared)
\begin{equation}
\frac{d\sigma_{T,L}}{dt}=\frac{1}{4\pi}\left\langle\left|
\int dz d^2x d^2y e^{iq_\perp.(z\textbf{x}\!+\!(1\!-\!z)\textbf{y})}
\Psi_{T,L}(z,\textbf{x}\!-\!\textbf{y})T_{\textbf{xy}}\right|^2\right\rangle_x\ ,
\label{diffcs}\end{equation}
where $2\Psi_T=\Psi_{V|\gamma}^{++}+\Psi_{V|\gamma}^{--}$ and $\Psi_L=\Psi_{V|\gamma}^{00}$ 
with
\begin{equation}
\Psi_{V|\gamma}^{\lambda'\lambda}(z,\textbf{r})=\sum_{h\bar{h}}
[\phi_{\lambda'}^{h\bar{h}}(z,\textbf{r})]^*\phi_{\lambda}^{h\bar{h}}(z,\textbf{r})\ ,
\label{overlap}\end{equation}
the overlap between the photon and meson wave functions. $\lambda$ and $h$ denote polarizations and helicities while $z$ is the longitudinal momentum fraction of the photon carried by the quark and
$\textbf{x}$ and $\textbf{y}$ are the quark and antiquark positions in the transverse plane.

The target average $\langle\ .\ \rangle_x$ is done with the CGC wave function squared
$|\Phi_x[{\cal A}^-]|^2:$
\begin{equation}
\langle f\rangle_x=\int DA^- |\Phi_x[A^-]|^2 f[A^-]\ .
\end{equation}
If one had imposed elastic scattering on the target side to describe the exclusive process
$\gamma^*A\!\to\!VA,$ the CGC average would be at the level of the amplitude, and the two-point
function $\langle T_{\textbf{xy}}\rangle_x$ inside the $|\ .\ |^2$ in (\ref{diffcs}), recovering the formula often used with dipole models.

\section{The McLerran-Venugopalan model}

Instead, when also including the target-dissociative part, the diffractive cross section
involves the 4-point correlator $\langle T_{\textbf{xy}}T_{\textbf{uv}}\rangle_x.$ In order to compute it, we must specify more about the CGC wave function. We shall use the McLerran-Venugopalan (MV) model \cite{mv}, which is a Gaussian distribution for the color charges which generate the field ${\cal A}:$
\begin{equation} |\Phi_x[A^-]|^2=\exp\left(-\int d^2xd^2y dz^+
\frac{\rho_c(z^+,\textbf{x})\rho_c(z^+,\textbf{y})}{2\mu^2(z^+)}\right)\ ,\label{MV}\end{equation}
where the color charge $\rho_c$ and the field ${\cal A}_c^-$ obey the Yang-Mills equation
$-\nabla^2{\cal A}_c^-(z^+,\textbf{x})=g_S\rho_c(z^+,\textbf{x}).$ The variance of the distribution is the transverse color charge density squared along the projectile's path $\mu^2(z^+),$ with
\begin{equation}
\langle\rho_c(z^+,\textbf{x})\rho_d(z'^+,\textbf{y})\rangle=\delta_{cd}\delta(z^+-z'^+)
\delta^{(2)}(\textbf{x}-\textbf{y})\mu^2(z^+)\ .
\end{equation}
The only parameter is the saturation momentum $Q_s,$ with $Q_s^2$ proportional to the integrated color density squared. Note that there is no $x$ dependence in the MV model, it should be considered as an initial condition to the small$-x$ evolution.

The MV distribution is a Gaussian distribution, therefore one can compute any target average by expanding the Wilson lines in powers of $g_S{\cal A}_c^-$ (see (\ref{wline})), and then use Wick's theorem \cite{fbgv}. All correlators of ${\cal A}$'s can be written in terms of
\begin{equation}
g_S^2\langle{\cal A}_c^-(x^+,\textbf{x}){\cal A}_d^-(y^+,\textbf{y})\rangle=\delta_{cd}\delta(x^+-y^+)g_S^4\mu^2(x^+)\int d^2z\ G(\textbf{x}-\textbf{z})G(\textbf{y}-\textbf{z})\ ,
\end{equation}
with the two-dimensional massless propagator
\begin{equation}
G(\textbf{x})=\!\int\limits_{|\textbf{k}|>\Lambda_{QCD}}\!\frac{d^2k}{{(2\pi)^2}}
\frac{e^{i\textbf{k}\cdot\textbf{x}}}{\textbf{k}^2}\ .
\end{equation}
The color algebra is the difficult part to deal with.

We give the result for the $\langle S_{\textbf{xy}}S_{\textbf{uv}}\rangle$ correlator \cite{dmw}, with
$S_{\textbf{xy}}=1-T_{\textbf{xy}}:$
\begin{eqnarray} \langle
S_{\textbf{xy}}S_{\textbf{uv}}\rangle=
\underbrace{e^{-\frac{C_F}2[F(\textbf{x}-\textbf{y})
+F(\textbf{u}-\textbf{v})]}}_{=\langle S_{\textbf{xy}}\rangle\langle S_{\textbf{uv}}\rangle}\
\left[\left(\frac{F(\textbf{x},\textbf{u};\textbf{y},\textbf{v})\!+\!\sqrt{\Delta}}{2\sqrt{\Delta}}
-\frac{F(\textbf{x},\textbf{y};\textbf{u},\textbf{v})}{N_c^2\sqrt{\Delta}}\right)
e^{\frac{N_c}4\mu^2\sqrt{\Delta}}\right.\nonumber\\\left.-
\left(\frac{F(\textbf{x},\textbf{u};\textbf{y},\textbf{v})\!-\!\sqrt{\Delta}}{2\sqrt{\Delta}}
-\frac{F(\textbf{x},\textbf{y};\textbf{u},\textbf{v})}{N_c^2\sqrt{\Delta}}\right)
e^{-\frac{N_c}4\mu^2\sqrt{\Delta}}\right]
\ e^{-\frac{N_c}4\mu^2F(\textbf{x},\textbf{u};\textbf{y},\textbf{v})
+\frac1{2N_c}\mu^2F(\textbf{x},\textbf{y};\textbf{u},\textbf{v})},\label{mainresult}\end{eqnarray}
where $\displaystyle\mu^2=\int dz^+\mu^2(z^+).$ This seemingly complicated object is given in terms of a single function:
\begin{equation}
F(\textbf{x}-\textbf{y})=g_s^4\mu^2\int\ d^2z
\left[G(\textbf{x}-\textbf{z})-G(\textbf{y}-\textbf{z})\right]^2\ .\end{equation}
Indeed also one has
\begin{eqnarray}
\Delta=F^2(\textbf{x},\textbf{u};\textbf{y},\textbf{v})+\frac4{N_c^2}
F(\textbf{x},\textbf{y};\textbf{u},\textbf{v})
F(\textbf{x},\textbf{v};\textbf{u},\textbf{y})\ ,\\
-2\mu^2F(\textbf{x},\textbf{y};\textbf{u},\textbf{v})
=F(\textbf{x}\!-\!\textbf{u})+F(\textbf{y}\!-\!\textbf{v})
-F(\textbf{x}\!-\!\textbf{v})-F(\textbf{y}\!-\!\textbf{u})\ . \end{eqnarray}

In the function $F(\textbf{r}),$ the infrared cutoff $\Lambda_{QCD}$ only enters through a logarithm as expected. In the $|\textbf{r}|\Lambda_{QCD}\!\ll\!1$ limit, one has
\begin{equation}
\frac{C_F}2F(\textbf{r})=\frac{g_S^4C_F}{2\pi}\mu^2
\int_{\Lambda_{QCD}}^\infty dk\ \frac{1-J_0(k|\textbf{r}|)}{k^3}\simeq\frac{\textbf{r}^2}4\ \underbrace{\frac{g_S^4C_F}{4\pi}\mu^2\log\left(\frac1{\textbf{r}^2\Lambda_{QCD}^2}\right)}_{\equiv Q_s^2(\textbf{r})}\ .
\label{qsatmv}\end{equation}
This is the standard definition of the saturation scale in the MV model. It is also possible to consistently include small$-x$ evolution in a such a calculation, in the large$-N_c$ limit
\cite{fgv} and beyond \cite{kkrw}. Essentially one should replace $\mu(z^+)$
by $\mu_x(z^+,\textbf{x},\textbf{y})$ in (\ref{MV}), and the $x$ evolution of the corresponding $F(x,\textbf{x},\textbf{y})$ can be obtained from the JIMWLK equation.

Finally, note that in (\ref{mainresult}) for $\textbf{x}=\textbf{y}$ or $\textbf{u}=\textbf{v},$
$F(\textbf{x},\textbf{y};\textbf{u},\textbf{v})=0$ and one recovers the single dipole average. Also, in the large$-N_c$ limit, one has
$\langle T_{\textbf{xy}}T_{\textbf{uv}}\rangle=\langle T_{\textbf{xy}}\rangle\langle T_{\textbf{uv}}\rangle,$ which means that at small$-x,$ the target-dissociative part of the diffractive cross-section in
suppressed at large $N_c,$ compared to the exclusive part.

\section{Results}

\begin{wrapfigure}{r}{0.45\columnwidth}
\vspace{-1.6cm}
\centerline{\includegraphics[width=0.4\columnwidth]{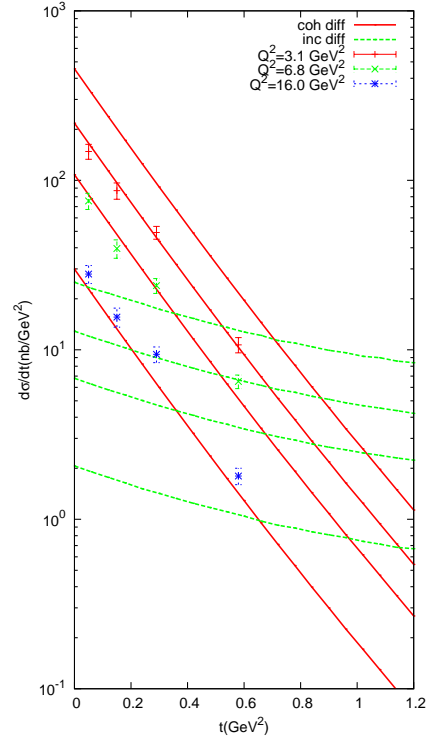}}
\caption{Diffractive J/$\Psi$ production in DIS at HERA, for $W=90\ \mbox{GeV}$ and different values of $Q^2.$ In our calculation, we separated the exclusive (full lines) and the proton-dissociative (dashed lines) parts.}
\end{wrapfigure}

The results presented in this section are obtained with the $x$ evolution of the saturation scale modeled as in \cite{gbw}:
\begin{equation}
Q_s(x)=\left(\frac{x_0}{x}\right)^{\lambda/2}\ \mbox{GeV}\ ,
\end{equation}
with $\lambda=0.277$ and $x_0=4.1\ 10^{-5}$ for the case of a target proton. The collinear logarithm of $Q_s$ (see (\ref{qsatmv})) is neglected, which corresponds to exact geometric scaling \cite{gs}: $F(x,\textbf{r})=F[\textbf{r}^2Q^2_s(x)].$ As an illustration, the resulting cross-section for diffractive J/$\Psi$ production is displayed in Fig.1, and separated into its coherent and incoherent contributions. The light-cone Gaussian J/$\Psi$ wave function \cite{jpsiwf} has been used in (\ref{overlap}). At small values of $|t|$ where coherent diffraction dominates, our results are in agreement with HERA data \cite{data} (one can get a better agreement with more realistic saturation models \cite{vm-sat}, but this is not our point). Our model indicates that for $|t|>0.7\ \mbox{GeV}^2$ or so (this value slightly decreases when $Q^2$ increases), incoherent diffraction starts to dominate. This may be the reason why the data on exclusive production stop: there is too much proton-dissociative `background'. We observe that this part of the cross-section decreases as a power law with $|t|,$ rather than exponentially as the exclusive part does.

In the case of a target nucleus, we expect the following qualitative changes in the $t$ dependence. First, the low$-|t|$ regime with elastic scattering of the nucleus will be dominant up to a smaller value of $|t|$ compared to the proton case, reflecting the bigger size of the nucleus. Then, the nucleus-dissociative part will be split into two: an intermediate regime in momentum transfer up to about $0.7\ \mbox{GeV}^2$ where the nucleus will predominantly break up into its constituents nucleons, and a large$-|t|$ regime where the nucleons inside the nucleus will also break up, implying pion production in the $Y$ system for instance.

The model discussed in this work is well adapted to describe the low- and large$-|t|$ regimes, but not the intermediate regime since the constituent nucleons are absent from the description
(\ref{cgc}). This problem has been addressed in a complementary setup in the case of inclusive diffraction off nuclei \cite{klmv}, and the coherent diffraction regime was found to be dominant up to about $|t|=0.05\ \mbox{GeV}^2.$ The vector meson production case will be addressed next. While in the proton case, both exclusive and diffractive processes can be measured, it is likely that at a future electron-ion collider, the exclusive cross section cannot be extracted: when the momentum transfer is small enough for the nucleus to stay intact, then it will escape too close to the beam to be detectable. Therefore the diffractive physics program will rely on our understanding of incoherent diffraction.


\begin{footnotesize}

\end{footnotesize}


\end{document}